\documentclass[preprint,prd,amsmath,amssymb,aps,nofootinbib]{revtex4-1}

\usepackage{graphicx}
\usepackage{bm}
\usepackage{natbib}
\def\mathbi#1{\textbf{\em #1}}

\def\apj{Astrophys.\ J.\ }
\def\apjl{Astrophys.\ J.\ Lett.\ }

\def\prd{Phys.\ Rev.\ D\ }

\def\apss{Astrophys.\ Space\ Sci.\ }

\begin{document}
\title{Test of the superdiffusion model in the interstellar medium around the 
Geminga pulsar}
\author{Sheng-Hao Wang$^{a,b}$}
\author{Kun Fang$^a$} \email{fangkun@ihep.ac.cn}
\author{Xiao-Jun Bi$^{a,b}$} \email{bixj@ihep.ac.cn}
\author{Peng-Fei Yin$^a$}

\affiliation{
 $^a$Key Laboratory of Particle Astrophysics, Institute of High Energy
 Physics, Chinese Academy of Science, Beijing 100049, China \\
 $^b$University of Chinese Academy of Sciences, Beijing 100049, China\\
 }

\date{\today}

\begin{abstract}
The TeV $\gamma$-ray halo around the Geminga pulsar is an important
indicator of cosmic-ray (CR) propagation in the local zone of the Galaxy as it
reveals the spatial distribution of the electrons and positrons escaping from
the pulsar. Considering the intricate magnetic field in the interstellar 
medium (ISM), it is proposed that superdiffusion model could be more realistic 
to describe the CR propagation than the commonly used normal diffusion model. 
In this work, we test the superdiffusion model in the ISM around the Geminga 
pulsar by fitting to the surface brightness profile of the Geminga halo 
measured by HAWC. Our results show that the chi-square statistic monotonously 
increases as $\alpha$ decreases from 2 to 1, where $\alpha$ is the 
characteristic index of superdiffusion describing the degree of fractality of 
the ISM and $\alpha=2$ corresponds to the normal diffusion model. We find that 
model with $\alpha<1.32$ (or $<1.4$, depending on the data used in fit) is 
disfavored at 95\% confidence level. Superdiffusion model with $\alpha$ close to 
2 can well explain the morphology of the Geminga halo, while it predicts 
much higher positron flux on the Earth than the normal diffusion model. This 
has important implication for the interpretation of the CR positron excess.
\end{abstract}

\maketitle

\section{Introduction}
\label{sec:intro}
Several middle-aged pulsars, such as the Geminga pulsar, are reported to be
surrounded by TeV $\gamma$-ray halos with scales larger than 20 pc
\cite{Abeysekara:2017old}. These $\gamma$-ray halos are generated by free
electrons and positrons\footnote{\textit{Electrons} will denote both electrons
and positrons hereafter.} diffusing out from the corresponding pulsar wind
nebulae (PWNe), rather than being interpreted by $\gamma$-ray PWNe
\cite{Giacinti:2019nbu}. According to the evolution model of PWN, the original
PWNe of the middle-aged pulsars were broken long time ago
\cite{Gaensler:2006ua}. These pulsars are currently traveling in the
interstellar medium\footnote{The pulsars could also be inside their old host
SNRs if the SNRs are large enough \cite{Kun:2019sks}.} (ISM) and driving
bow-shock PWNe with scales $\lesssim1$ pc. This is consistent with the
observations of the X-ray PWN of the Geminga pulsar
\cite{2003Sci...301.1345C,2006ApJ...643.1146P,Posselt:2016lot}. The TeV
$\gamma$-ray halos are generated mainly through the inverse Compton scattering
(ICS) of the homogeneous cosmic microwave background, so the morphologies
of the halos unambiguously indicate the electron propagation in the ISM.

Cosmic-ray (CR) propagation in the ISM is usually modeled by the
diffusion process considering the turbulent nature of the ISM
\cite{1964ocr..book.....G}. The diffusion approximation is based on the
assumption that the inhomogeneity of the chaotic magnetic field has small-scale
character and is negligible in terms of the scale of interest. However,
multi-scale inhomogeneities could exist in the ISM. The ISM is more likely
to be a fractal type and the normal diffusion can be generalized to
superdiffusion 
\cite{2001NuPhS..97..267L,2003NIMPB.201..212L,2012GrCo...18..122U,
2017arXiv170306486U}, where the CR propagation is simulated by L\'{e}vy
flights instead of the Brownian motion.  This model has been
applied in the Galactic-scale propagation of CRs to explain features of
the CR energy spectra
\cite{2001NuPhS..97..267L,2015JPhCS.632a2027V}. In this work, we test the
superdiffusion model in a local region of the Galaxy by
explaining the morphology of the Geminga halo. As the diffusion packets are
different for the normal diffusion and superdiffusion models, the $\gamma$-ray
morphologies predicted by them could also be distinct.

This paper is organized as follows. In Section~\ref{sec:method}, we introduce 
the superdiffusion model and the solution of the propagation equation. Then we 
briefly introduce the information of Geminga in Section~\ref{sec:geminga}. We 
fit the $\gamma$-ray surface brightness profile (SBP) of the Geminga halo 
measured by HAWC with different propagation models and discuss the fitting 
results in Section~\ref{sec:result}. In Section~\ref{sec:posi} we discuss the 
impact on the interpretation of the positron excess according to the results of 
Section~\ref{sec:result}. Finally, we conclude in Section~\ref{sec:conclude}.

\section{The Superdiffusion model}
\label{sec:method}
After escaping from the source, CRs are continuously scattered by the chaotic
magnetic field in the ISM. If the chaotic magnetic field is uniformly
distributed, the particle transportation can be simulated by the Brownian
motion \cite{1964ocr..book.....G}. However, the realistic magnetic field in the ISM may
consist of turbulent and relatively regular components, where very long
jumps for CRs are permitted. L\'{e}vy flight, which is characterized by the
occasionally very long steps, should be the more proper description in this
case
\cite{2001NuPhS..97..267L,2003NIMPB.201..212L,2012GrCo...18..122U,
2017arXiv170306486U}. For one-dimensional L\'{e}vy flight, the probability
density function (PDF) of the individual step length $x$ has the heavy-tailed
form of $P(x)\propto|x|^{-1-\alpha}$ for $x\rightarrow\infty$, where
$0<\alpha<2$. Obviously, the variance of the step length is infinite, which is
different from that of the Brownian motion. The widening of the diffusion
packet with time is also faster for L\'{e}vy flight ($\propto t^{1/\alpha}$)
than the normal diffusion case ($\propto t^{1/2}$), so this scenario is named
after superdiffusion or accelerated diffusion
\cite{2012GrCo...18..122U}.

As the PDF of L\'{e}vy flight can be described by the fractional Laplacian
equation, the propagation equation of CR electrons should be
\begin{equation}
  \frac{\partial N(E, \mathbi{r}, t)}{\partial t} = -D(E, \alpha)
(-\Delta)^{-\frac{\alpha}{2}} N(E, \mathbi{r}, t) +
\frac{\partial[b(E)N(E, \mathbi{r}, t)]}{\partial E} + Q(E,
\mathbi{r}, t)\,,
 \label{eq:prop}
\end{equation}
where $N$ is the differential number density of electrons, $D$ is the diffusion
coefficient, and $Q$ is the source function for which we will present the
details in Section~\ref{sec:geminga}. The index $\alpha$ represents the
degree of fractality of the ISM and for $\alpha=2$ the equation degenerates to
the normal diffusion case. The radiative energy-loss rate $b(E)=b_0E^2$ induced 
by synchrotron radiation and ICS must be considered for high-energy electrons. 
The magnetic field strength is assumed to be 3 $\mu$G to obtain the synchrotron 
term. For the ICS process, we adopt the seed photon field in 
Ref.~\cite{Abeysekara:2017old} and the parameterization given by
Ref.~\cite{2020arXiv200715601F} to calculate the energy-loss rate, where the 
Klein-Nishina correction is accurately considered.

\begin{figure}[t]
 \centering
 \includegraphics[width=0.65\textwidth]{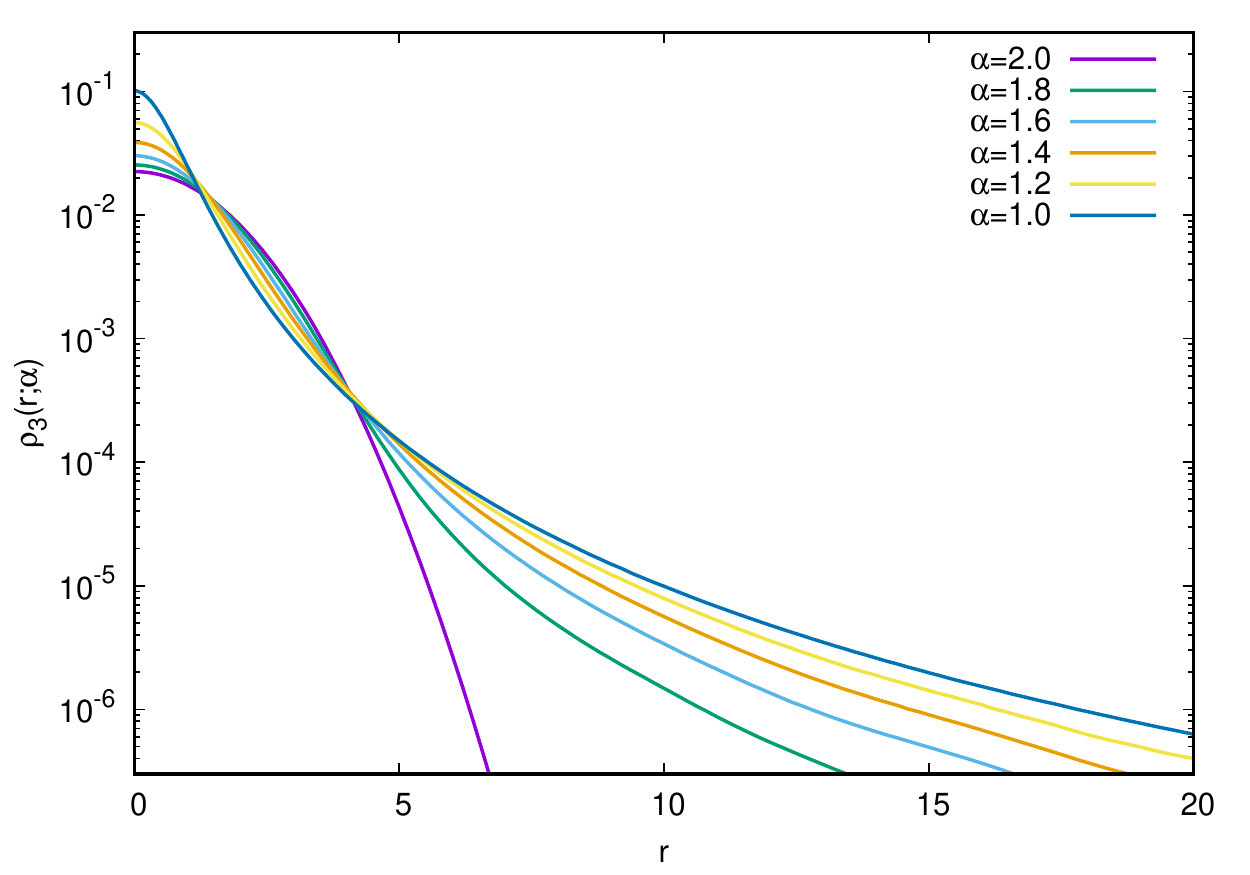}
 \caption{Probability density functions of three-dimentional spherically
symmetric stable distributions for different $\alpha$.}
 \label{fig:rho}
\end{figure}

We solve Eq.~(\ref{eq:prop}) with the Green function method. The Green function
takes the form of
\begin{equation}
 G(E, \mathbi{r}, t; E_0, \mathbi{r}_0,
t_0)=\frac{\rho^{(\alpha)}_3(|\mathbi{r}-\mathbi{r}_0|\lambda^{-1/\alpha})}{
b(E)\lambda^{3/\alpha}}\delta(t-t_0-\tau)H(\tau)\,,
 \label{eq:green}
\end{equation}
where
\begin{equation}
\tau=\int_E^{E_0}\frac{dE'}{b(E')}\,,\quad \lambda=\int_E^{E_0}\frac{D(\alpha,
E')}{b(E')}dE'\,,
 \label{eq:tau}
\end{equation}
and $H$ is the Heaviside step function. In Eq.~(\ref{eq:green})
$\rho_3^{(\alpha)}(r)$ is the PDF of a three-dimensional spherically-symmetrical
stable distribution with index $\alpha$. The exact expression of
$\rho_3^{(\alpha)}(r)$ is
\begin{equation}
  \rho_3^{(\alpha)}(r) = \frac{1}{2{\pi^2}r} \int_0^\infty{e^{k^{\alpha}}
\sin(kr)kdk}\,,
 \label{eq:rho}
\end{equation}
while in practice it can be expressed with convergent and asymptotic series
\cite{1999cs..book.....U}:
\begin{equation}
  \rho_3^{(\alpha)}(r) = \frac{1}{2\pi^2
\alpha}\sum_{n=0}^{\infty}{\frac{(-1)^n}{(2n+1)!}\Gamma(\frac{2n+3}{\alpha})r^{
2n}}\,,
 \label{eq:rho2}
\end{equation}
\begin{equation}
  \rho_3^{(\alpha)}(r) = \frac{1}{2\pi^2
r}\sum_{n=1}^{\infty}{\frac{(-1)^{n-1}}{n!}\Gamma(n\alpha +2)\sin(\frac{n\alpha
\pi}{2})r^{-n\alpha -2}}\,.
 \label{eq:rho3}
\end{equation}
We show $\rho_3^{(\alpha)}(r)$ in Fig.~\ref{fig:rho}, where we can clearly see
the difference between the heavy-tailed distributions for $\alpha<2$ and the
Gaussian distribution for $\alpha=2$. The solution of Eq.~(\ref{eq:prop}) can
be then expressed as
\begin{equation}
 \begin{aligned}
 N(E, \mathbi{r}, t) = & \int_{R^3}
d^3\mathbi{r}_0\int_{-\infty}^{t}dt_0\int_{-\infty}^{+\infty}
dE_0\,G(E, \mathbi{r}, t; E_0, \mathbi{r}_0, t_0)\,Q(E_0, \mathbi{r}_0, t_0) \\
 = & \int_{R^3} d^3\mathbi{r}_0\int_{t-1/(b_0E)}^{t}dt_0\,
\frac{b(E_\star)}{b(E)}\frac{\rho^{(\alpha)}_3(|\mathbi{r}-\mathbi{r}
_0|\lambda^{ -1/\alpha})}{\lambda^{3/\alpha}}\,Q(E_\star,
\mathbi{r}_0, t_0)\,,
 \end{aligned}
 \label{eq:solution}
\end{equation}
where $E_\star\simeq E/[1-b_0E(t-t_0)]$.

From Eq.~(\ref{eq:solution}), we can get the electron number density at 
arbitrary distance from a pulsar. To calculate the $\gamma$-ray SBP around the
pulsar, we integrate the electron number density along the line-of-sight from 
the Earth to the vicinity of the pulsar and get the electron surface density at 
an arbitrary angular distance $\theta$ from the pulsar:
\begin{equation}
 F(\theta)=\int_0^{+\infty}N(l_\theta)dl_\theta\,,
 \label{eq:los}
\end{equation}
where $l_\theta$ is the length in that line-of-sight, and $N(l_\theta)$ is the
electron number density at a distance of
$\sqrt{d^2+l_\theta^2-2dl_\theta\cos\theta}$ from the pulsar, where $d$ is the
distance between the pulsar and the Earth. With $F(\theta)$ and the standard
calculation of ICS \cite{Blumenthal:1970gc}, we can finally obtain the
$\gamma$-ray SBP around the pulsar.

\section{The Source Geminga}
\label{sec:geminga}
The Geminga halo is so far the best-studied TeV halo. The $\gamma$-ray SBP is
precisely measured by HAWC \cite{Abeysekara:2017old} and is an ideal object
to investigate the propagation of CR electrons.
In the normal diffusion scenario, the derived diffusion coefficient around
the Geminga pulsar is more than two orders of magnitude smaller than the
average value in the Galaxy indicated by the boron-to-carbon ratio
\cite{Aguilar:2016vqr}. The origin of the slow diffusion has been discussed in
recent works \cite{Evoli:2018aza,Kun:2019sks,Liu:2019zyj}.

The parameters of the Geminga pulsar can be obtained from the ATNF catalog
\cite{Manchester:2004bp}. The age of the pulsar is $t_s=342$ kyr and the
current spin-down luminosity is $L=3.25\times10^{34}$ erg s$^{-1}$. The latest
version of the catalog gives the pulsar distance as $d=190$ pc and provides the
reference Ref.~\cite{1996ApJ...461L..91C}, where the distance is derived
with the optical measurement of the trigonometric parallax. However, the
distance given by Ref.~\cite{1996ApJ...461L..91C} is 157 pc rather than 190 pc.
Here we adopt $d=250$ pc which is determined by the latest parallax measurement
\cite{2007Ap&SS.308..225F}.

Electrons are accelerated to very high energy inside the Geminga PWN. As the
scale of the Geminga PWN is significantly smaller than the TeV halo, it is
reasonable to assume it to be a point source. The time dependency of the
electron injection is assumed to be proportional to the spin-down luminosity of
the pulsar as $\propto(1+t/t_{\rm sd})^{-2}$, where $t_{\rm sd}$ is the
spin-down time scale of pulsar. We set a typical value of $t_{\rm sd}=10$ kyr. 
As the cooling time of 100 TeV electrons is about 10 kyr, the parent electrons 
of the present TeV $\gamma$-ray halo are generated in the very recent age of 
Geminga. So the value of $t_{\rm sd}$ impacts little on the time profile as 
$t/t_{\rm sd}\gg1$. The injection energy spectrum can be described by a
power-law within the energy range of interest as $\propto E^{-p}$. We adopt
$p=2.24$ which is measured by the HAWC work \cite{Abeysekara:2017old}. Thus,
the source term can be written as
\begin{equation}
 Q(E,\mathbi{r},t)=\left\{
 \begin{aligned}
 & Q_0(E/E_{100})^{-p}\,\delta(\mathbi{r}-\mathbi{r}_s)\,[(t_s+t_ { \rm
sd})/(t+t_{\rm sd})]^2\,, & t>0 \\
 & 0\,, & t<0
 \end{aligned}
 \right.\,,
 \label{eq:src}
\end{equation}
where $\mathbi{r}_s$ is the position of the pulsar and we set the birth time of
the pulsar to be the zero point of time. The constant $Q_0$ is the current-time
normalization at the energy of $E_{100}=100$ TeV.

\section{Fitting to the HAWC data}
\label{sec:result}
We test the superdiffusion propagation in the regime of $1\leq\alpha\leq2$ by
fitting to the $\gamma$-ray SBP of HAWC \cite{Abeysekara:2017old}. For each
$\alpha$, we seek the best-fit model by minimizing the chi-square statistic
$\chi^2$ between the model and the data points.
The NLopt\footnote{http://github.com/stevengj/nlopt} package and the
optimization algorithm BOBYQA \cite{bobyqa} are adopted for the fitting
procedures. The free parameters are the anomalous diffusion coefficient $D$ and
the constant factor of the source term $Q_0$. Since HAWC provides the SBP in a
single energy bin of 8-40 TeV which is not very broad, we assume an
energy-independent $D$ in the calculations.

HAWC measures the SBP within 10$^\circ$ around the Geminga pulsar. However, we
do not use all the SBP data in the fitting procedures as the data at large
angular distances could be affected by other potential $\gamma$-ray sources. As
indicated by Fig.~\ref{fig:rho}, the differences of particle distribution
between the propagation models are still significant at large distances from
the source. If we include the data at large angular distances, the $\chi^2$
test may be disturbed by the inhomogeneous $\gamma$-ray background. We use the
data within $\theta_{\rm max}=4^\circ$ ($\approx17$ pc) around the pulsar as a
benchmark and test the case of $\theta_{\rm max}=6^\circ$ ($\approx26$ pc) for
comparison.

\begin{figure}[t]
 \centering
 \includegraphics[width=0.7\textwidth]{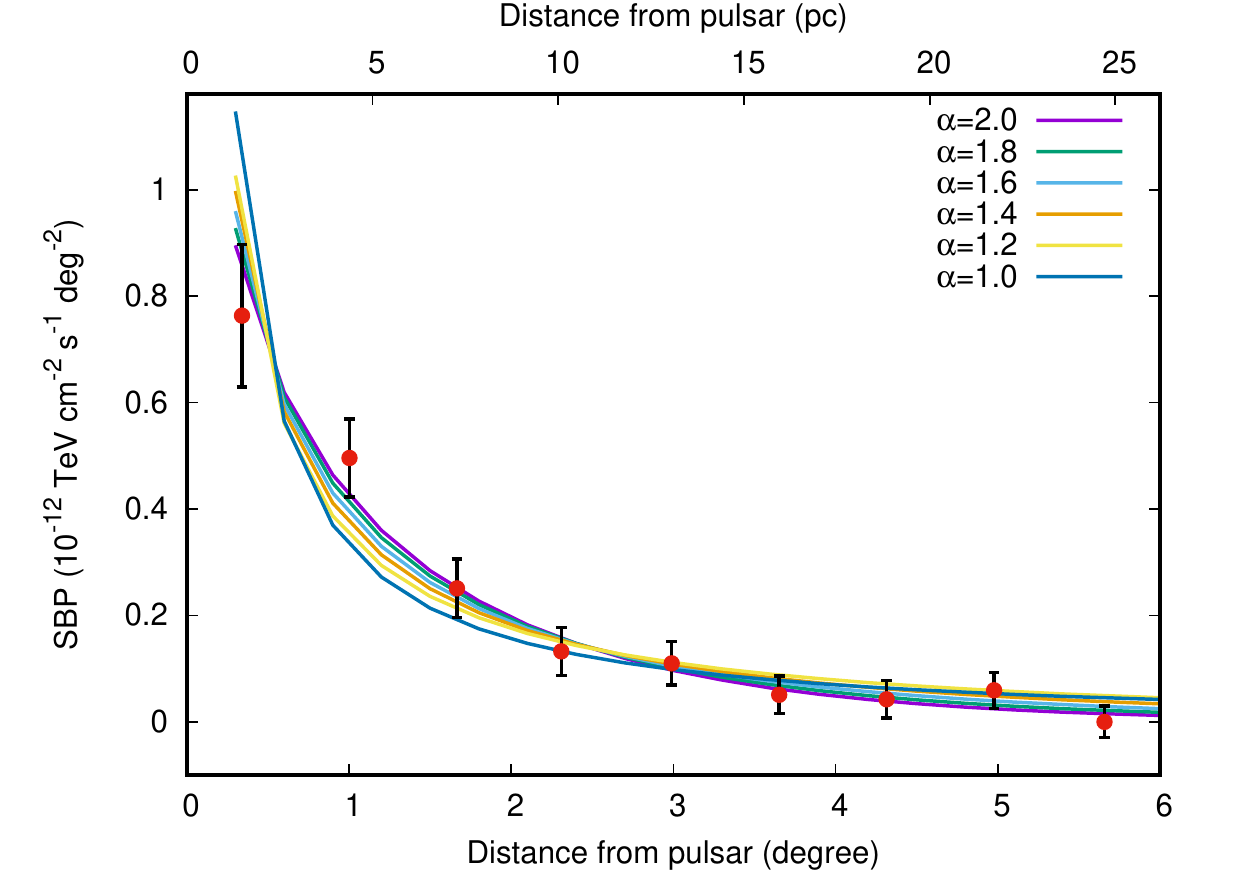}
 \caption{Best-fit surface brightness profiles to the HAWC data
\cite{Abeysekara:2017old} with the normal diffusion model ($\alpha=2$) and 
superdiffusion models ($\alpha<2$).}
 \label{fig:profile}
\end{figure}

\begin{table}[t]
 \centering
 \caption{Best-fit parameters towards the HAWC data (using the data within
4$^\circ$ around the Geminga pulsar).}
 \begin{tabular}{ccccccc}
  \hline
  \hline
  $\alpha$ & 2.0 & 1.8 & 1.6 & 1.4 & 1.2 & 1.0 \\
  \hline
  $\log_{10}[D({\rm cm^\alpha~s^{-1}})]$ & ~27.4~ & ~23.6~ & ~19.9~ & ~16.3~ &
~12.7~ & ~9.1~~\\
  \hline
  $Q_0(10^{28}~{\rm TeV^{-1}~s^{-1}})$ & 2.5 & 2.9 & 3.5 & 4.8 & 9.2 & 15.6 \\
  \hline
 \end{tabular}
 \label{tab:param}
\end{table}

\begin{figure}[t]
 \centering
 \includegraphics[width=0.48\textwidth]{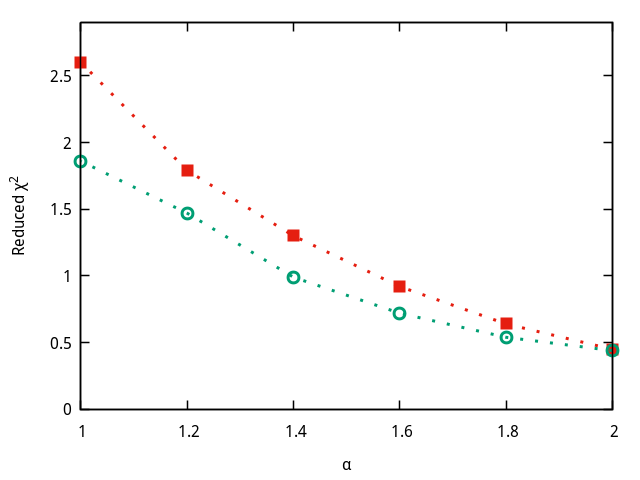}
  \includegraphics[width=0.48\textwidth]{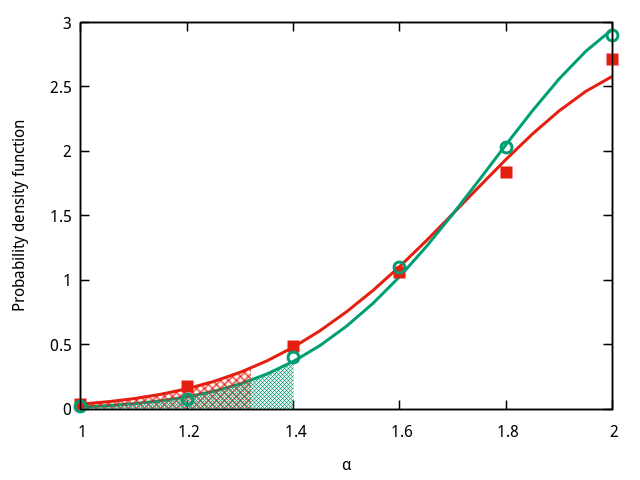}
 \caption{Left: Reduced $\chi^2$ of the fitting procedures for different
assumptions of $\alpha$. The red filled squares correspond to the results in
Fig.~\ref{fig:profile} where the HAWC data within 4$^\circ$ around the pulsar
are used. The case using the data within 6$^\circ$ is shown with the green
empty circles. Right: The corresponding probability density functions of
$\alpha$ derived with $\chi^2(\alpha)$. The shaded areas show the intervals
at 5\% CL.}
 \label{fig:chi2}
\end{figure}

The fitting results are presented in Fig.~\ref{fig:profile} and the best-fit
parameters are listed in Table \ref{tab:param}. Superdiffusion models with 
$\alpha\lesssim1.5$ yield too steep inner profiles of the $\gamma$-ray
halo as can been seen in Fig.~\ref{fig:profile}. Due to the nature of L\'{e}vy
flight, the superdiffusion models predict steeper profiles close to the source
and flatter profiles far from the source in comparison with the normal
diffusion case. If the inner fluxes are forced to fit the data for the
superdiffusion models, the outer fluxes will be much higher than the data. The
reduced $\chi^2$ ($\chi^2$ divides degree of freedom) monotonously increases as
$\alpha$ decreases from 2 to 1, which is shown in the left panel of
Fig.~\ref{fig:chi2}. The results indicate that the normal diffusion model is
still the best depiction among the diffusion scenarios in terms of the current
measurement.

We then provide a quantitative constraint on $\alpha$ by assuming it as a
variable parameter and deriving its one-dimensional distribution. According to 
the Bayesian inference, we have $P(\alpha)\propto
L(\alpha)\propto{\rm exp}[-\chi^2(\alpha)/2]$, where $P(\alpha)$ is the
posterior PDF and $L(\alpha)$ is the likelihood function. The relative
probability densities can be derived at the knots where $\chi^2(\alpha)$ are
available. We then create a PDF with the form of
\begin{equation}
P(\alpha)=c_0\left\{{\rm
exp}\left[-\frac{(\alpha-c_1)^2}{2c_2^2}\right]-{\rm
exp}\left(-\frac{c_1^2}{2c_2^2}\right)\right\}
\end{equation}
to fit the relative probability densities, where $c_0$, $c_1$, and $c_2$ are
free parameters. The expression ensures $P(0)=0$ since the domain of $\alpha$
is (0,2] \cite{2008IJBC...18.2649D}. We determine the free parameters by the 
least square method and then rescale $c_0$ to satisfy the normalization 
condition $\int_0^2P(\alpha)d\alpha=1$. The obtained PDFs are shown in the right 
panel of Fig.~\ref{fig:chi2}. The points in the figure are the relative 
probability densities used for the fits, which are rescaled after the 
normalizing of $P(\alpha)$. The best-fit parameters are $c_0=2.69$, $c_1=2.11$, 
and $c_2=0.383$ for the case of $\theta_{\rm max}=4^\circ$ and $c_0=3.13$, 
$c_1=2.12$, and $c_2=0.349$ for the case of $\theta_{\rm max}=6^\circ$. Finally, 
we exclude $\alpha<1.32$ for the former case and $\alpha<1.40$ for the latter at 
95\% confidence level (CL).

\section{Impact on the positron excess}
\label{sec:posi}
The Geminga pulsar was considered as one of the most competitive candidate
sources of the CR positron excess 
\cite{Hooper:2008kg,Yuksel:2008rf,Yin:2013vaa}. However, if the slow diffusion
measured by HAWC pervades the ISM between Geminga and the solar system, Geminga
can hardly contribute to the positron flux at the Earth as the positrons do not
have enough time to reach the Earth \cite{Abeysekara:2017old}. This problem is 
alleviated if the slow diffusion only happens in the nearby ISM of the Geminga 
pulsar \cite{Hooper:2017gtd,Fang:2018qco,Profumo:2018fmz,Tang:2018wyr}. On the 
other hand, it is possible that the diffusion coefficient on the Galactic disk 
is generally much smaller than the average value in the Galaxy
\cite{2016PhRvD..94l3007F,2018PhRvD..97f3008G}.

\begin{figure}[t]
 \centering
 \includegraphics[width=0.48\textwidth]{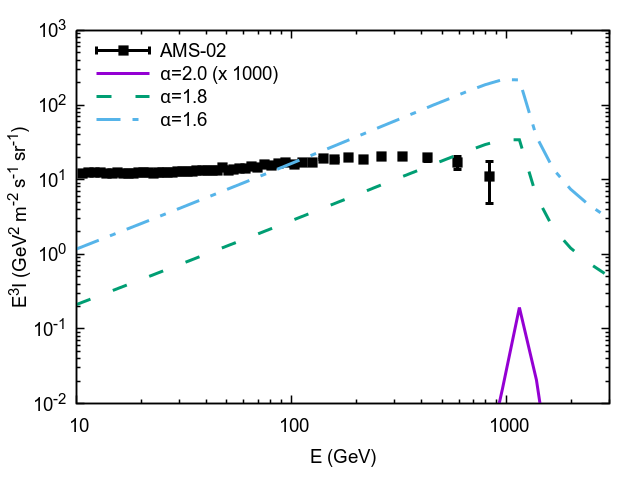}
 \includegraphics[width=0.48\textwidth]{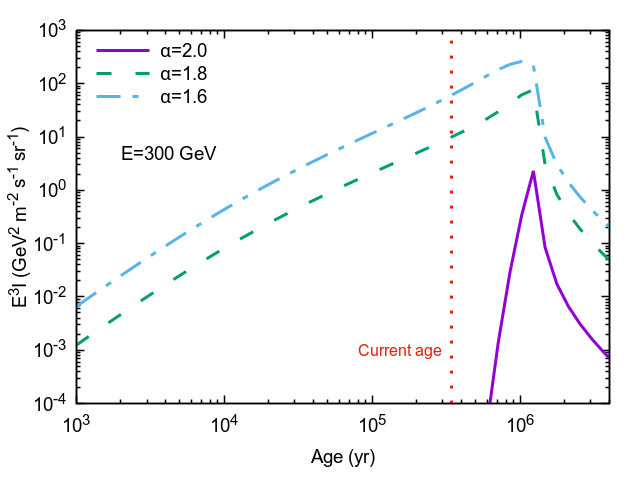}
 \caption{Left: Positron spectra on the Earth contributed by Geminga with both
the normal diffusion model and superdiffusion models. The parameters are
adopted from Tabel \ref{tab:param}, which are derived from the HAWC
measurement of the Geminga halo. The AMS-02 positron spectrum is also shown for
comparison \cite{Aguilar:2019owu}. Right: Positron fluxes from Geminga at 300
GeV as functions of the pulsar age, corresponding to the models in the left
panel.}
 \label{fig:elec}
\end{figure}

The discussion in the last paragraph is based on the normal diffusion model.
The results in Section~\ref{sec:result} indicate that superdiffusion model with
$\alpha>1.4$ is permitted in terms of the current measurement. As can be
seen in Fig.~\ref{fig:rho}, superdiffusion models yield much higher fluxes in
large distance from the source than the normal diffusion due to the
heavy-tailed distribution. We assume a one-zone superdiffusion scenario to
calculate the positron flux from Geminga. The left panel of Fig.~\ref{fig:elec}
shows the positron spectra in the cases of $\alpha=2.0$, $\alpha=1.8$, and
$\alpha=1.6$ adopting the parameters in Table \ref{tab:param}. The diffusion 
coefficient is extrapolated from the value in the HAWC energy range with the 
relation of $D\propto E^{1/3}$, which is predicted by the Kolmogorov's theory. 
As expected, Geminga can contribute significant positron flux in comparison 
with the AMS-02 data \cite{Aguilar:2019owu} in the superdiffusion models, 
without the assumption of two-zone propagation mentioned above. More 
intuitively, we show the positron fluxes from Geminga at 300 GeV as functions of 
the pulsar age in the right panel of Fig.~\ref{fig:elec}. Positrons come much 
faster to the Earth in the superdiffusion models with power-law-like time 
dependencies, rather than the steep time dependency of $\sim{\rm 
exp}[-r^2/(4Dt)]$ for the normal diffusion model. The flux cutoff above 
$10^6$~yr is due to the radiative cooling of 300~GeV positrons.

Note we still adopt $p=2.24$ in the above calculations, which are derived with
the TeV observation of HAWC. As the AMS-02 positron spectrum is in the GeV
range, it is more reasonable to use the Fermi-LAT observation of Geminga to
constrain the injection spectrum and diffusion coefficient in the same energy
range \cite{Shao-Qiang:2018zla}. However, the constraint from Fermi-LAT is
model-dependent and the analysis is beyond the scope of this work.

\section{Conclusion}
\label{sec:conclude}
In this work, we test the superdiffusion model in the ISM
around the Geminga pulsar by fitting to the SBP measured by HAWC. This model
depicts the particle propagation in a fractal medium with L\'{e}vy flight,
which could be more realistic to simulate the intricate magnetic field
environment compared with the normal diffusion. The L\'{e}vy flight
superdiffusion is described by the fractional Laplacian in the propagation
equation with the order of $\alpha/2$, where $\alpha\in(0,2)$. Through the
$\chi^2$ test, we find that the normal diffusion ($\alpha=2$) still give the
best accommodation to the HAWC data. The reduced $\chi^2$ monotonously
increases with the decrease of $\alpha$ and $\alpha<1.32$ (or $\alpha<1.4$,
depending on the data used in fit) is disfavored at 95\% CL. Intuitively, 
models with small $\alpha$ give poor fits to the $\gamma$-ray fluxes close to 
the pulsar. With more TeV halos being accurately measured in the coming future, 
particle propagation in the local zones of the Galaxy can be further 
constrained.

Superdiffusion model with $\alpha>1.4$ is still permitted in terms of the
current measurement of the Geminga halo. Models with $\alpha$ close to 2 can
give comparable fitting results to that of the normal diffusion model,
however, they predict distinct positron spectra on the Earth. Due to the nature
of the heavy-tailed distribution of the superdiffusion model, part of the
positrons come much faster to the Earth and the positron flux from Geminga is
much higher than that predicted by the normal diffusion model.
Different from the conclusion in Ref.~\cite{Abeysekara:2017old}, Geminga could
have significant contribution to the observed high-energy positron spectrum 
in the superdiffusion scenario even if the small diffusion coefficient
measured around the Geminga pulsar is applied in the whole region between
Geminga and the Earth.

The test may also provide information for the origin of the inefficient 
particle propagation in TeV halos. For example, it is proposed that the 
slow-diffusion zone around the Geminga pulsar may be interpreted by its crushed
relic PWN \cite{Profumo:2018fmz,Tang:2018wyr}. Considering the filamentary
structures in the relic PWN due to Rayleigh-Taylor instabilities
\cite{Blondin:2001bf}, this scenario may be described by the superdiffusion 
model with $\alpha$ significantly smaller than 2, which can then be tested by 
the method of this work. However, quantitative relations between the 
superdiffusion model and specific physical scenarios need to be established in 
the future.

\acknowledgments
This work is supported by the National Key Program for Research
and Development (No.~2016YFA0400200) and by the National Natural Science
Foundation of China under Grants No.~U1738209,~11851303.

\bibliography{/home/fangkun/work/paper/ref_ins/references}

\end{document}